\documentstyle[aps,prl]{revtex} 
\tighten
\begin{document}
\input epsf.sty
\flushbottom
\draft
\twocolumn[\hsize\textwidth\columnwidth\hsize\csname @twocolumnfalse\endcsname 
\title{
Are Quantum-Cascade Lasers really quantum?
}
\author{Rita Claudia Iotti$^{1,2}$ and Fausto Rossi$^{1}$}
\address{
$^{1}$Istituto Nazionale per la Fisica della Materia (INFM) and 
Dipartimento di Fisica,\\ 
Politecnico di Torino, Corso Duca degli Abruzzi 24, 10129 Torino, Italy \\
$^{2}$Scuola Normale Superiore, Piazza dei Cavalieri 7, 56126 Pisa, Italy
}

\date{\today}
\maketitle

\begin{abstract}

The first {\it global simulation} of semiconductor-based quantum-cascade 
lasers is presented; Our fully three-dimensional approach allows to study 
in a purely microscopic way ---without resorting to phenomenological 
pa\-ra\-me\-ters--- the current-voltage characteristics of 
state-of-the-art unipolar nanostructures.
Based on the proposed theoretical scheme, we are able to give a definite
answer to the long-standing controversial question: 
{\it is charge transport in quantum-cascade lasers mainly coherent or 
incoherent?}
Our analysis clearly shows that a proper inclusion of carrier-phonon as 
well as carrier-carrier scattering within a semiclassical framework 
gives excellent agreement with experimental results. 

\end{abstract}

]

\narrowtext

Since the seminal paper of Esaki and Tsu~\cite{Esaki}, semiconductor-based 
nanometric heterostructures have been the subject of an impressive 
theoretical and experimental activity, due to their high potential 
impact in both fundamental and applied research~\cite{B-R,T-A}.
One of the main fields of research focuses on exploiting ``band-gap 
engineering'', namely the splitting of the bulk conduction band into 
several subbands, to generate and detect 
electromagnetic radiation in the infrared spectral region, as already 
envisioned by Kazarinov and Suris~\cite{KS}.

Unipolar coherent-light sources like quantum-cascade lasers 
(QCLs)~\cite{QCL94}, are usually modelled in terms of three-levels 
systems~\cite{QCL}. Their theoretical description is thus often grounded 
on purely macroscopic models: the only relevant physical quantities are 
the various carrier concentrations within the different energy levels, 
whose time evolution is dictated by proper sets of coupled rate equations.
This simplified three-level picture does not take into account the
existence of transverse or ``in-plane'' degrees of freedom and thus provides 
no information on the microscopic processes governing carrier dynamics within 
the three-dimensional (3D) multi-subband structure.
As pointed out in~\cite{APL}, such a macroscopic modeling can only operate 
as an {\it a posteriori} fitting procedure: The best-fit phenomenological 
parameters obtained through a comparison with experiments do strongly depend 
on details of the 3D hot-carrier distribution and therefore on the device 
operation conditions.

QCLs are complex devices, whose core is a multi-quantum-well (MQW) 
structure made up of repeated stages of active regions sandwiched between
electron-injecting and collecting regions. When a proper bias is applied,
an electron cascade along the subsequent quantized-level energy staircase 
takes place.  
For a detailed understanding of the basic physical processes involved,
two main issues need to be addressed: 
(i) the nature of the hot-carrier relaxation within the device active region; 
(ii) the nature ---coherent versus incoherent--- of the physical mechanisms
governing charge transport through injector/active-region/collector 
interfaces.

Point (i) has been recently addressed in \cite{APL}, where the macroscopic 
treatment of hot-carrier relaxation within the device active region 
previously mentioned has been compared to a fully kinetic description, 
based on the following set of equations~\cite{APL}:
\begin{equation}
{d \over dt} f_{{\bf k}\nu} =
\left[
g_{{\bf k}\nu} - \Gamma_{{\bf k}\nu} f_{{\bf k}\nu}
\right]_{{\rm i/l}} 
+ \! \sum_{{\bf k}'\nu'} \! \left[
P_{{\bf k}\nu,{\bf k}'\nu'} f_{{\bf k}'\nu'} 
- P_{{\bf k}'\nu',{\bf k}\nu} f_{{\bf k}\nu} 
\right]\ . 
\label{KE1}
\end{equation}
Here, the first two terms describe ---still on a partially phenomenological
level--- injection/loss (i/l) of carriers with wavevector ${\bf k}$ in 
subband $\nu$, while the last ones describe intra- as well as inter-subband 
in- and out-scattering processes (${\bf k}\nu \to {\bf k}'\nu'$) within the 
device active region only.
The Boltzmann-like structure of (\ref{KE1}) allows for a partially stochastic 
solution; to this aim, the commonly used Monte Carlo (MC) method \cite{MC} 
has been adopted. 
The simulated experiments in \cite{APL} show that, as far as the active 
region is concerned, the quantum-cascade process is mainly determined by 
carrier-optical phonon emission; Moreover, the values of effective 
interlevel rates strongly depend on the shape of the in-plane charge 
distributions as well as on Pauli-blocking effects.

However, the microscopic analysis in \cite{APL}, being limited to the device 
active region only, does not allow to answer point (ii): 
which are the basic physical mechanisms governing charge transport through 
injector/active-region/collector interfaces?
This issue is intimately related to the long-standing controversial 
question~\cite{RT}: 
{\it is charge transport in quantum-cascade lasers mainly coherent or 
incoherent?}

To provide a definite answer to this fundamental question, in this Letter 
we present the first {\it global simulation} ---injector plus active region 
plus collector--- of semiconductor-based QCL structures.
To this end, the partially phenomenologic model in (\ref{KE1}) has to be 
replaced by a fully microscopic description of the whole MQW core structure. 
This is not a trivial task. It requires: 
(i) a proper description of all the 3D electron states within the active as 
well as injector/collector regions; 
(ii) a fully microscopic simulation scheme to ``close the circuit''. 

As a starting point, we evaluate ---within a self-consistent 
Schr\"odinger-Poisson approach in an envelope-function 
framework~\cite{APL,Barbieri99}--- the set of 3D single-particle electron 
states $\{{\bf k}\nu\}$ corresponding to a single QCL stage, e.g., active 
region plus collector/injector in the presence of the applied bias 
(see Fig.~\ref{fig1}).
Given such carrier states, we consider the ideal MQW structure obtained 
as infinite repetition of this QCL periodicity region (see Fig.~\ref{fig1}).
Within such extended scheme, the time evolution of the carrier distribution 
function $f_{{\bf k}\alpha}$ is governed by the following Boltzmann-like 
equation:
\begin{equation}
{d \over dt} f_{{\bf k}\alpha} =
\sum_{{\bf k}'\alpha'} \! 
\sum_s
\left[
P^s_{{\bf k}\alpha,{\bf k}'\alpha'} f_{{\bf k}'\alpha'} 
- P^s_{{\bf k}'\alpha',{\bf k}\alpha} f_{{\bf k}\alpha} 
\right] \ .
\label{KE2}
\end{equation}
Here, $\alpha \equiv (\lambda,\nu)$ denotes the generic electron state in 
our MQW structure, i.e., the $\nu$-th state of the $\lambda$-th stage,
while $s$ labels the different scattering mechanisms considered, e.g., 
carrier-phonon, carrier-carrier, etc.

To ``close the circuit'', we impose periodic boundary conditions limiting 
the inter-stage ($\lambda' \ne \lambda$) scattering to just nearest-neighbor 
coupling ($\lambda' = \lambda \pm 1$).
In view of the translational symmetry, we are allowed to simulate carrier 
transport over the central ---i.e., $\lambda = 0$--- stage only. 
Every time a carrier in state $\nu$ undergoes an inter-stage scattering 
process (i.e., $0,\nu \to \pm 1,\nu'$, the electron is properly reinjected 
into the central region ($0,\nu \to 0,\nu'$) and the corresponding 
electron charge $\pm e$ will contribute to the current through the 
device~\cite{periodicity}.
The proposed global-simulation scheme allows to study in a purely 
microscopic way ---without resorting to phenomenological i/l parameters--- 
the current-voltage characteristics of state-of-the-art unipolar 
nanostructures.

At this point a few comments are in order.
Contrary to Eq.~(\ref{KE1}), the Boltzmann-like equation in (\ref{KE2}) 
---describing an infinite system--- does not contain explicit i/l terms, 
which are automatically accounted for within the proposed re-injection scheme.
Moreover, the electric field does not enter our transport equation via the 
usual drift term \cite{Barbieri99}; the corresponding bias is directly taken 
into account in evaluating the single-particle states 
$\{{\bf k}\nu\}$~\cite{periodicity}. 

It is well-known that the application of the present semiclassical transport 
theory to MQW structures at high fields becomes questionable. 
Indeed, according to (\ref{KE2}), in the absence of scattering processes 
($P_{{\bf k}\alpha,{\bf k}'\alpha'} = 0$) we have no current. 
This paradoxical result is due to the neglect of phase coherence between 
our localized states. A proper treatment of coherent transport 
---including resonant-tunneling effects--- would require a density-matrix 
formalism \cite{DMA}.
However, in the presence of a strong scattering dynamics current is mainly 
determined by incoherent hopping processes between localized states; 
moreover, such phase-breaking processes lead to a very fast decay of 
off-diagonal density-matrix elements. In this regime coherent-transport 
effects play a minor role and the semiclassical limit is recovered.
It is then clear that the nature ---coherent versus incoherent--- of charge
transport in QCLs is primarily dictated by the role played by incoherent 
scattering processes.

In order to discriminate between the two different regimes, we have 
applied the above global-simulation scheme to state-of-the-art QCL 
structures.
In particular, as prototypical device, we have considered the 
diagonal-configuration QCL in \cite{Sirtori98}, schematically depicted 
in Fig.~\ref{fig1}, in which the proposed simulation scheme is also sketched. 
Here, the energy levels and probability densities of various electron states 
within the simulated stage ($\lambda = 0$) are also plotted: 
They correspond to 
the device active region (full line, $\nu = 1,2,3$ according to the standard 
notation) 
as well as to the collector region (dashed line, $\nu =$ A,B,C,D,E). 
The nearest stages $\lambda = \pm 1$ are partially shown for clarity as 
well.

In order to properly model phase-breaking hopping processes, in addition 
to carrier-optical phonon scattering, all various intra- as well as 
intersubband carrier-carrier interaction mechanisms have been considered. 
To this end, the well-established MC method for the simulation of 
intercarrier scattering in quantum-well systems~\cite{MC} has been 
extended to our MQW biased structure. Within such approach 
---based on the Born approximation--- 
the effect of nonequilibrium multisubband screening 
of the bare two-body Coulomb potential is taken into account within the
standard random-phase approximation~\cite{GL}.
Other scattering mechanisms, not included in the simulation, are expected to 
play a very minor role~\cite{MC}: The interaction with acoustic phonons 
---which in the absence of carrier-carrier scattering would be essential 
to get ergodicity--- does not affect energy relaxation significantly due to
its quasi-elastic nature and small coupling constant; Carrier-impurity 
scattering is also expected to have very little impact on 
vertical-transport phenomena.

As a first step, we have investigated the relative weigth of the 
carrier-carrier and carrier-phonon competing energy-relaxation channels.
The time evolution of the carrier population in the various subbands as 
well as of the total current density in the
presence of carrier-phonon scattering are depicted in Fig.~2.
Parts (a) and (b) refer, respectively, to simulated experiments without and 
with two-body carrier-carrier scattering.
In our ``charge conservation'' scheme, we start the simulation assuming the 
total number of carriers to be equally distributed among the different 
subbands; then the electron distribution functions evolve according to 
Eq. (\ref{KE2}) and a steady-state condition is eventually reached, 
leading to the desired $3 \rightarrow 2$ population inversion.
As we can see, the inclusion of intercarrier scattering has significant 
effects;
It strongly increases inter-subband carrier redistribution, thus reducing 
the electron accumulation in the lowest energy level A and 
optimizing the coupling between active region and injector/collector 
(the populations of subbands 3 and B get equal).
This effect comes out to be crucial in determining the electron flux 
through the MQW structure. 
Indeed, Fig.~2(c) shows the simulated current density across the core 
region of our QCL device, obtained with (full line) and without (dashed line) 
the inclusion of carrier-carrier scattering.
These simulations have been performed at the threshold operating parameters
estimated in~\cite{Sirtori98}. The current density in the presence of both 
electron-phonon and electron-electron scattering mechanisms is about 4 
kA/cm$^{2}$: This value is in good agreement~\cite{agreement} 
with the experimental results in ~\cite{Sirtori98}.  

The results plotted in fig.~\ref{fig2} clearly demonstrate that the 
electron-phonon interaction alone is not able to efficiently couple the 
injector subbands to the active region ones: 
While carrier-phonon relaxation well describes 
the electronic quantum cascade within the bare active region~\cite{APL}, 
carrier-carrier scattering plays an essential role in determining 
charge transport through the full core region.
This can be ascribed to two typical features of carrier-carrier interaction 
---compared to the case of carrier-phonon---:  
(i) this is a long-range two-body interaction mechanism, which also couples 
non-overlapping single-particle states [see Fig.~1];
(ii) the corresponding scattering process at 
relatively low carrier density is quasielastic, thus coupling nearly 
resonant energy levels, like states 3 and B.

This result is extremely important: it tells us that the presence of 
a coherent-tunneling dynamics ---often invoked to account for nearly 
resonant inter-level processes between active region and 
injector/collector--- is not necessary: 
The charge transport in QCL structures can be explained quantitatively in 
terms of incoherent intercarrier-assisted tunneling processes, 
and thus within a purely semiclassical transport theory.
This allows us to answer the long-standing controversial question on the 
nature ---coherent versus incoherent--- of charge transport in QCLs 
previously mentioned: 
for the typical structures considered, energy-relaxation and dephasing 
processes are so strong to destroy any phase-coherence effect; as a result,
the usual semiclassical description is in excellent agreement with 
experimental results.

Since in the proposed MC simulation scheme, the current density across 
the whole structure is an output of the simulation, the 
current-versus-voltage 
behavior of our semiconductor quantum device can be obtained
within our fully microscopic description, without resorting to any 
phenomenological external parameter. 
Figure 3 presents the simulated voltage-current characteristics for the
MQW structure of Fig.~\ref{fig1}. 
As we can see, the current density increases with the applied field, due to
the increased coupling between subband 3 and various lower-energy injector 
states, in which carriers tend to relax and accumulate. 
The agreement between our theoretical results and the measured dependence 
shown in Ref.~\cite{Sirtori98} is extremely good. 
Discrepancies between simulated and measured values are well within the
measurement uncertainties and the complexity of the real device structure.

In summary, we have presented the first global simulation of 
semiconductor-based QCLs. The proposed theoretical scheme allows for a
fully microscopic evaluation of the current-voltage characteristics of 
prototypical QCL structures. 
The excellent agreement between our semiclassical simulations and recent 
experimental results confirms the incoherent nature of charge transport in 
such unipolar quantum devices.

\medskip\par
This work has been supported in part by the European Commission through the 
Brite Euram UNISEL project (Contract No. BE97-4072). 
\begin{figure}
\caption{\label{fig1} 
Schematic representation of the conduction-band profile along the growth
direction for the diagonal-configuration QCL structure of 
Ref.~\protect\cite{Sirtori98}. 
The MQW is biased by an electric field of 48 kV/cm.
The levels $\nu = 1,2,3$ and $\nu =$ A,B,C,D,E in the active and 
collector regions (full and dashed lines, respectively) of the simulated 
stage ($\lambda = 0$) are also plotted together with the corresponding 
probability densities. 
The replica of level 3 in the following stage $\lambda = +1$ is shown for 
clarity.
}
\end{figure}
\begin{figure}
\caption{\label{fig2} 
Time evolution of simulated carrier densities in the various subbands
of the MQW structure of Fig.~\protect\ref{fig1} (full lines: 
$\nu =$ A,B,C,D,E; dotted lines: $\nu =$ 1,2,3), without (a) and with (b) 
intercarrier scattering. 
(c): current density across the whole structure in presence (full line) 
and absence (dotted line) of carrier-carrier interaction. 
An electric field of 48 kV/cm has been applied.
}
\end{figure}
\begin{figure}
\caption{\label{fig3} 
Applied-field vs current-density characteristic at 77K of the MQW 
structure of Fig.~\protect\ref{fig1}. 
The dashed line is a guide to the eye.
}
\end{figure}

\begin{references}
\bibitem{QCL}
The basic idea is to identify a three-level quantum structure: 
carriers are injected into the upper level (level 3) which is 
``metastable'', i.e., weakly coupled to the lower levels; in contrast, 
carriers in the intermediate level (level 2) can relax very fast into the 
lowest level (level 1).
As a result, the desired population inversion between levels 3 and 2
is obtained.
\bibitem{periodicity}
Due to the system periodicity, the energy levels $\epsilon_\alpha$ 
along the growth direction are of the form:
$\epsilon_\alpha = \epsilon_{\lambda,\nu} = 
\epsilon_{0,\nu} + \lambda V_0$, $V_0$ being the 
potential drop per period. Thus, every time a carrier is reinjected into 
the simulation stage, its total energy is changed by $\pm V_0$.
\bibitem{agreement}
The apparent discrepancy between the experimental ($\simeq$ 7 kA/cm$^2$) 
and the theoretical values ($\simeq$ 4 kA/cm$^2$) is due, we believe, 
to the different estimate of the potential drop per period required to 
line up the ground state A of the injector with level 3 of the active region. 
Our single particle calculation, which takes into account Poisson 
renormalization of the band structure, allows to predict a good alignment
at a field of 68 kV/cm. This corresponds to a potential drop of 9.2 V 
along the device thirty-period core region, in much better agreement with 
the measured value of 11.4 V than the 6.9 V at a field of 48 kV/cm quoted 
in Ref.~\protect\cite{Sirtori98}.
As shown in Fig.~\protect\ref{fig2}(c) the simulated current density at
68 kV/cm is about 7 kA/cm$^2$, thus enforcing the present interpretation.
\end{references}
\end{document}